\documentclass[11pt,review,1p,a4paper]{elsarticle}
\biboptions{sort&compress}
\usepackage{longtable}
\usepackage{graphicx}     
\usepackage{color}
\usepackage{hyperref}
\usepackage{amsmath}
\usepackage{url}
\usepackage{natbib}

\newcommand{\etal}{{\it et al.}}
\newcommand{\ai}{{\it ab~initio}}

\newcommand{\cm}{cm$^{-1}$}

\newcommand{\ka}{$K_a$}
\newcommand{\red}[1]{{\color{red} #1}}

\def\a0{{$a_{\rm 0}$}}

\begin{document}

\title{Variational analysis of HF dimer tunneling rotational spectra
using an {\it ab initio} potential energy surface}

\author[UCL,NN]{Oleg L. Polyansky{\footnote{To whom correspondence  should be addressed; email: o.polyansky@ucl.ac.uk}}}
\author[NN]{Roman I. Ovsyannikov}
\author[UCL]{Jonathan Tennyson}
\author[NN]{Sergei P. Belov}
\author[NN]{Mikhail {Yu}. Tretyakov}
\author[NN]{Vladimir Yu. Makhnev}
\author[NN]{Nikolai F. Zobov}

\address[UCL]{Department of Physics and Astronomy, University College London, Gower Street, London WC1E 6BT, United Kingdom}
\address[NN]{Institute of Applied Physics, Russian Academy of Sciences, 46 Ulyanov Street, Nizhny Novgorod, 603950, Russia}
\date{\today}

\begin{abstract}                                  

A very accurate, (HF)$_2$  potential energy surface (PES)
due to Huang \etal ({\it J. Chem. Phys.}, {\bf 150}, 154302 (2019)) is used to
calculate  the energy levels of the HF dimer by solving the nuclear-motion
Schr\"{o}dinger equation using variational program WAVR4.
%(Kozin \etal, Computer Phys. Comm., {\bf 100}, 2323, (2004)).
Calculations on an extended range of rotational states show very good agreement
with experimental data. In particular 
the known empirical rotational constants for
 the ground and some observed excited
vibrational states  are reproduced with an accuracy of about 50 MHz.
This level of accuracy is shown to extend to  higher excited inter-molecular 
vibrational states $v$ and higher excited rotational quantum numbers $(J,K_a)$. 
These calculations allow the assignment of 3 new $J$-branches 
in an HF dimer tunneling-rotation spectra recorded 
30 years ago. These branches belong to 
excited \ka\ = 4  state of the ground vibrational state, and
 \ka\ = 0 states of excited inter-molecular vibrational states.
\end{abstract}
\maketitle

\section{Introduction}
 The study of weakly bound molecular complexes with hydrogen bonds  and Van der Waals bonding accounts
 for a large sub-branch of 
 molecular spectroscopy \cite{88Nexxxx.HFHF}. 
 High resolution spectroscopy can provide very accurate information on these
 complicated quantum systems but often requires theoretical support to provide interpretation and understanding. 
 High precision study of hydrogen bonded species also
 extends beyond molecular physics to the study of
 biological systems, while still relying on a quantum
 mechanical interpretation. 
 
  Quantum chemical calculations of the HF dimer potential energy
  surface (PES) started in 1974 \cite{74YaNeSc.HFHF}. Already the 1991
 review by Quack and Suhm \cite{91QuSuxx.HFHF} cites 35 papers reporting \ai\ PESs.  
 Further work on the \ai\ (HF)$_2$ PESs  is 
 presented, for example, by Klopper, Quack and Suhm \cite{96KlWiQu.HFHF,98KlWiQu.HFHF}. 
 The PES used here is due to Huang \etal \cite{19HuYaZh.HFHF} and is based on 
 about 100~000 points calculated using a very high level of theory. The accuracy
 of this PES is demonstrated below in Section \ref{sec:calcs}. 
  
  Most molecular complexes can only be observed experimentally  in cold jets or
 molecular beams. In the gas phase at room temperature these complexes dissociate into
  monomers or  condense into polymers 
 on the gas cell walls.
  The HF dimer is one of the rare 
  molecular complexes  possessing a large dimerisation 
 constant which therefore results in sufficient collisionally-formed complexes at pressures appropriate for high 
 resolution spectroscopy at higher temperatures than achievable in 
 molecular jets and beams. Other such examples include HF-H$_2$O~\cite{07BeDeZo.H2O-HF} and 
  HF-HCN~\cite{87LeMiWi.HCN-HF}.
% and some
% others, which can be observed in the gas phase. . 
 %and several others.
 %This advantage 
 %of such complexes gives us ability to observe their spectra in the highly excited
 %vibrational and rotational states.
 A major reason for
 studying such complexes is that thermal occupation allows one to observe spectra involving highly excited
 vibrational and rotational states.
 That is the main reason why HF dimer spectra are so well studied~\cite{96QuSuxx.HFHF}. 
 On the other hand these gas phase observations highlight 
 significant incompleteness in the theory of HF dimers. As discussed below, observations made of HF dimer spectra
 in highly excited vibrational states remain unassigned 
 due to the limitations of theory.

This paper presents a significant step forward in the accurate representation
of highly excited energy levels of the HF dimer. 
We present first principles variational calculations
of  rotationally-excited HF dimer states  and compare them with the high $J$
energies, derived from line frequencies
obtained from high resolution HF dimer spectra. Moreover, we use these calculations
to provide first assignments to  lines in the tunneling-rotation-vibration spectra
of the HF dimer, recorded 30 years ago \cite{90BeKaKo.HFHF}.  
At the time of our recording this sub-millimeter spectrum of
HF dimer at equilibrium in the gas phase, the theoretical 
tools
for assigning high resolution HF dimer spectra were rather limited.
 The microwave spectrum of the HF  dimer was first observed
by Dyke \etal ~\cite{72DyHoKl.HFHF}, and then Pine and Lafferty studied 
its mid infrared (IR) spectrum  in the region of the HF monomer fundamental stretch 
\cite{83PiLaxx.HFHF}. Subsequently, a thorough 
study of the gas
phase HF dimer  millimeter region spectrum was 
performed by Lafferty \etal\ \cite{87LaSuLo.HFHF}. The spectrum of the HF dimer 
behaves in many respects like that of a
 linear molecule, as for certain vibrational states and 
$K$ values  the rotation-tunneling spectrum consists of the characteristic quasi-equidistant branches with increasing
values of the total rotational  quantum number $J$. This behavior can   be 
described by a simple formula (see Eq. (1) of Section II).
So although the HF dimer is an asymmetric top and, 
in principle, the asymmetric top quantum number \ka\ 
should be used,  it is actually almost a symmetric top and one can  use
the designation $K$  instead. 

%\cite{Howard72,Lafferty}
When in 1990 we recorded the spectrum of the 
HF dimer in what at the time was the  more exotic and difficult to work in sub-millimeter 
wave region 
around 10  \cm\ \cite{90BeKaKo.HFHF},
the first obvious thing to do was to
continue the branches observed in the lower part of the
millimeter region \cite{87LaSuLo.HFHF}
to higher
frequencies and, as we dealt  mostly with the R-branches, to
higher $J$s of the same $K$ rotational quantum numbers.
We did that and assigned the $K=0,1$ and 2 R-branches  
of the tunneling-rotation spectrum. 
As the HF dimer is a floppy, quasi-linear system, the extrapolation to higher $K$'s, first of all to $K=3$,
proved impossible using the standard effective Watson Hamiltonian. Fortunately,
by that time the method of one-dimensional approximation of effective
Hamiltonians in a Pad\'{e}-Borel form had 
already been developed by one of us \cite{85Poxxxx.H2O} and it was applied
to the extrapolation to higher $K$'s of the ground and first excited 
bending state of water monomer \cite{87BeKoPo.H2O,96PoBuGu.H2O}.
Before that approximants in Pad\'{e} form were used by 
Pine \etal\ \cite{84PiLaHo.HFHF}
to assist assignment of the $K=4$ branch in the $\nu_1$ band of (HF)$_2$.
The use of approximants allowed us to extrapolate \cite{90BeKaKo.HFHF} 
the $B$ constants for $K=0,1$ and 2 to the  
$B$ constant
for $K=3$. Using this prediction we identified the $K=3$ branch of the  
tunneling
rotation spectrum. However, since the Q-branch of this spectrum  was in
the millimeter region, beyond the capability of the spectrometer we used to record
the spectrum, we had to ask our colleagues at the National Institute of Standards 
and Technology (NIST) to observe the $K=3$ Q-branch
we predicted, as well  the continuation of our R-branch
series of lines. Our colleagues at NIST
successfully observed the Q-branch  using our predictions. Thus, our
extrapolation from $K=0, 1$ and 2 to $K=3$ was confirmed as being correct and the assignment of
the $K=3$ branch established \cite{90BeKaKo.HFHF}.

Nevertheless,  at that time we were unable to go further. We observed
many unassigned lines belonging to the HF dimer, but could not do anything with them because
of limitations in the effective Hamiltonian
technique, even in its more sophisticated Pad\'{e}-Borel form. Although the
Pad\'{e} approximation efficiently sums the divergent perturbation series
of the effective Hamiltonian and extrapolates well to higher rotational
quantum numbers, it can do nothing with the vibrational spectrum;
the majority of unassigned lines in our spectrum were
assumed to be due to the vibration-tunneling-rotation transitions
in four low-lying inter-molecular vibrational states $\nu_3$,
$\nu_4$, $\nu_5$ and $\nu_6$. 
The values of the $J=0$ energy levels of these bands are respectively
about 475  \cm\ \red{\cite{98KlWiQu.HFHF}}, 125 \cm\ \cite{98KlWiQu.HFHF}, 160 \cm\ \cite{98KlWiQu.HFHF} and 370 \cm\ \cite{91QuSuxx.HFHF}. These values are
comparable with the ground state values of
 116 \cm\ (for $K=J=2$) and  235 \cm\ (for $K=J=3$). It means
that the Boltzmann factor of the low $K$ ($K=0$ and 1)  lines in these excited 
vibrational states is
comparable to the Boltzmann factor of the excited $K$ lines in the ground
vibrational state, so we should have been able to see them. This suggests that
many of the observed unassigned lines  belonged to these excited vibrational
states and not, for example, to  impurities. 

Variational calculations on the HF dimer with $J=0$ have a long history. 
Most \ai\ PESs have been produced by Suhm, Quack and their co-workers, see Refs. \cite{91QuSuxx.HFHF,96KlWiQu.HFHF,98KlWiQu.HFHF,
95QuSuxx.HFHF,90QuSuxx.HFHF}
and references therein.
These were
%suhm1 -Quack M, Suhm M,Mol.Phys.,v.69, 791(1990)
%suhm2 JCP,v.95,28(1991), suhm3 - Chem.Phys.Lett,v.234,71(1995)
%suhm4 Klopper W, Quack M, Suhm M,ChemPhysLett,v.261,35(1996)
%suhm5 JCP,v.108,10096(1998)
full dimensional 6D calculations. There were also reduced dimensional (frozen monomer) calculations
by Bunker and Jensen \cite{90JeBuKa.HFHF} and a (4,2)D calculation by Quack, Suhm 
and others \cite{96KlWiQu.HFHF}. 
So far all full 6D
variational calculations only deal with $J=0$ states;  highly excited
$J$s have been studied only using less accurate Quantum Monte Carlo (QMC)
calculations \cite{99WuHaMc.HFHF}. The $J=0$ calculations 
of Quack and Suhm augmented by their QMC
excited $K$ calculations helped  
them to assign many spectra in the $\nu_5$ and $\nu_6$ bands in the IR region \cite{91QuSuxx.HFHF}.
However, in the
tunneling rotation sub-millimeter region, where all the HF dimer lines belonging to various vibrational states
are mixed together, $J=0$ only calculations are not enough. 

In this paper we  present variational calculations on (HF)$_2$ for the excited $J$
states; these are described in detail in the next section.
Section \ref{sec:experiment} summarizes the 1990 experimental
sub-millimeter spectrum of HF dimer.
Section \ref{sec:analysis} describes the application of our calculations to the
analysis of this spectrum. The first information on tunneling-rotational high resolution spectra in the $v_4 = 1$ and $v_3 = 1$ states
is given. We also assigned the  $K = 4$ branch of 
rotation-tunneling spectrum in the 
 ground vibrational state.
Section \ref{sec:concl} presents our conclusions.

\section{Variational calculations of HF dimer tunneling rotation-vibration energy levels}
\label{sec:calcs}

Variational calculations of excited rovibrational
tunneling energy levels of the HF dimer are performed using the \ai\ PES calculated  
by Huang \etal~\cite{19HuYaZh.HFHF}.  This PES is both of a high theoretical level,
using CCSD(T) with an aug-cc-pvQZ basis set, and built on a very large number of calculated 
\ai\ points -- 100~000, which makes the resulting surface very smooth.
Use of this gives  highly accurate predictions of 
rovibrational-tunneling 
energy levels. In particular, the discrepancies between 
calculated and empirical values (Tables
\ref{tab:obs.semi.vib} and \ref{tab:obs.Ks} below) are at least as accurate as the best empirical ($i.e.$ spectroscopically
determined or, in other words, adjusted to experimental data) PESs.
%\red{fitted is not so clear in this context. If you mean fitted to spectra, I would say best spectroscopically
%determined rather just fitted. After all one fits \ai\ PESs too.} PESs.
 The HF dimer empirical PES
of Klopper, Quack and Suhm \cite{98KlWiQu.HFHF} reproduced the available
experimental data on $K=0$ and excited $K$ energy levels better than any of the then available \ai\ 
surfaces.  This work used QMC
calculations of  excited rotational $J$ and $K$ levels to fit the  (HF)$_2$ PES to experimental energy levels with $K>0$  \cite{98KlWiQu.HFHF}.  However, QMC calculations  
of nuclear motion energy levels lack accuracy; for example, this work only computed
rotational $B$ constants  with 
1 \% accuracy. We are unaware of any previous 
variational calculations of highly excited $J$ and $K$ states; such calculations are necessary to obtain
accurate energy levels beyond  band origins. These energies are needed
for many purposes, in particular
to provide the more numerous theoretical counterparts to experimental
measurements for accurate characterisation of the PES and 
also to assist assignment of the spectra.

%However, our goal is to calculate  HF dimer energy levels  as accurately as possible \ai, without empirical fitting
%to the experimental energy levels, and with the PES containing as many  \ai\ calculated points as possible. 

For the variational solution of the nuclear motion problem we 
use  the WAVR4 \cite{jt339}
program suite, developed especially for the calculation of energy
levels and wavefunctions  of   tetratomic
molecules with large amplitude internal motions. 
WAVR4  uses an exact, body-fixed
kinetic energy operator \cite{jt312,jt346}. The program offers a choice of
orthogonal (or polyspherical) coordinates. For %this MT: "this" was used above to designate the work by Klopper et al.,
the present work we 
used diatom -- diatom
Jacobi coordinates and discrete variable representation (DVRs) based on
Morse-oscillator-like functions to represent the three radial coordinates.
Rotationally excited states are treated using a 
two-step procedure which involves
solving an initial set of effective vibrational problems which excludes Coriolis coupling
and then using the results as a basis to solve the full problem \cite{jt46}.  
The original implementation
of this procedure in WAVR4 \cite{jt479} proved to be rather inefficient for higher
rotational states \cite{jt553} but has  been improved so that for the present
calculations the computer time taken only depends linearly on $J$. 
This new version of WAVR4, which includes the capability of 
computing dipole transition intensities, is currently being prepared for publication \cite{newWAVR4}.

%In the literature we did not find
%any variational calculations of excited $J$ levels.
Calculations were performed for $J$ up to 9. This allows us to calculate both $B$, the rotational constant, and $D$,  the distortion constant,
for corresponding branches of the HF dimer spectrum as well as many 
theoretical line frequencies directly (up to $J=9$) without resorting to
 simplified treatments of rotational excitation, using formulae of the type
\begin{equation}
\label{eq:simple}
E =        BJ(J+1) - D J^2(J+1)^2 .
\end{equation}
However, energy levels with  $J > 9$, if needed, 
could be obtained using this formula.

Before describing our excited $J$ calculations, we consider 
the $J=0$ calculations. These calculations provide the first step in the high $J$ calculations 
and also illustrate the accuracy with which  our adopted PES 
reproduces the empirical-determined frequencies of 
all four inter-molecular vibrations.

  After the matching of the PES \cite{19HuYaZh.HFHF} with the program
  WAVR4, by transformation of coordinate systems and the units used, 
  calculation of the vibration-rotation levels of the HF dimer  
consisted of the following steps:
\begin{enumerate}
%\item Connecting 
%capacity\cite{19HuYaZh.HFHF} to the 
%program for calculating vibrational-rotational energy 
%levels of four-atomic molecules WAVR4. 
\item Choice  of an integration region  
where the potential is well defined, 
namely, for the HF monomer stretches $1.48\ a_0 <r_1, r_2 <2.4\ a_0$ and for the HF -- HF stretch $3.5\ a_0 <R <40\ a_0$ (where $r_1$ and $r_2$ are internuclear distances of the two HF monomers% 1 and 2
, and $R$ is the distance between centers of mass of the monomers). 
To select a mesh extending to $R > 20$~a$_0$, it 
was necessary to modify WAVR4 by increasing the number 
of iterations in the LAGUER algorithm which selects the location 
and weights of integration points. The LAGUER algorithm in WAVR4 is an improved version of a LAGUER algorithm from
DVR3D \cite{jt338}. We used the following Morse-like oscillator
parameters which are designed to give DVR grids over the appropriate integration range:
for the monomer stretches we used
$r_e$ = 1.03~\AA, $w_e$ = 7500 \cm , $D_e$ = 43000 \cm\  and 
for the inter-molecular stretch   $R_e$ = 7.7~\AA,  
$w_e$ = 16 \cm , $D_e$ = 5000 \cm. 
\item To ensure convergence, a sufficiently large basis must be used to represent the 
vibrational-rotational states of interest. Only 
10 DVR points were needed for the stiff monomer $r_1$ and $r_2$ coordinates, 
while the inter-molecular stretch required 90 DVR points to obtain a good
representation. For bending modes, $l_{max} = 16$ basis functions 
were used, and $k_{max}$ = 6 basis functions for internal rotation, where $l_{max}$ determines sizes of H$_1$F$_1$F$_2$ and H$_2$F$_2$F$_1$ bending basis sets 
and $k_{max}$ gives the  size of H$_1$F$_1$-H$_2$F$_2$ internal rotation basis set. 
The quality of convergence becomes apparent from the shift in energy 
levels with increasing basis parameters. Thus, with an increase in 
$l_{max}$ to 20, the energies up to 500 \cm\  changed by no more than 
$10^{-4}$ \cm . With an increase in $k_{max}$ to 10, the energies up to 
500 \cm\ changed by no more than 10$^{-3}$ \cm . 
Increasing the DVR grid  along $r_1$ and $r_2$ to 16 points (with a change in 
the basis parameter $w_e = 12000$  \cm\ to maintain the same integration 
interval), the energies of fundamental inter-molecular 
vibrations changed by no more than 1 \cm.
%up to 500 \cm\ changed by no more than 2.5 \cm . 
While increasing the DVR grid along $R$ to 
100 points (with a change in the basis parameter $w_e = 18$ \cm\  
to maintain the same 
integration interval), 
%the energies up to 500 \cm\  changed by no more 
the energies of fundamental inter-molecular 
vibrations changed by no more than 0.5 \cm.
A final vibrational Hamiltonian matrix of dimension 7200 was used. This size comes
from truncation at 80 basis function for each of the 90 radial points. 
Increasing the truncation from 80 up to 120 basis functions leads to a test 
vibrational Hamiltonian matrix of dimension 10800, and shows the following convergence 
of the main Hamiltonian: $6 \times 10^{-5}$ \cm\ for energies below 500 \cm\ 
and $2\times 10^{-4}$ \cm\ for energies below 1000 \cm. 
\end{enumerate}

Table \ref{tab:obs.semi.vib} presents a comparison of our  $J=0$ calculations with  
values obtained empirically. Drawings of all six HF dimer fundamental vibrational
modes are given, for example, in Fig.~2 of the paper by Quack \etal\  \cite{91QuSuxx.HFHF}.  
There are no accurate high resolution data on the frequencies of  these modes.
However, there have been several attempts to obtain relatively accurate indirect
estimates of their values. In particular, Nesbitt and co-authors \cite{96AnDaNeb.HFHF,96AnDaNea.HFHF} used high resolution
combination band spectra to estimate the fundamental frequencies. 
In Table I of their paper \cite{96AnDaNea.HFHF}
%Nesbit2 - Anderson DT, JCP, v.105, 4498, 1996
%Nesbit1 - JCP, v.104, 6225 (1996)
%MT: either $\nu_4$ or $v_4=1$
$\nu_4$ and $\nu_5$ fundamental frequencies are given  with an estimated
accuracy of 0.1 \cm. For $\nu_4$ the frequency was given as 125 \cm,
whereas for $\nu_5$ it was estimated as 160 \cm.
These two frequencies derived from extrapolation of 
high resolution spectra of combination bands  $\nu_1+\nu_4$, 
$\nu_2+\nu_4$,
$\nu_1+\nu_5$ and $\nu_2+\nu_5$. Less accurate extrapolations have been performed for the
higher energy inter-molecular modes $\nu_3$ and $\nu_6$. Table II
of paper \cite{96AnDaNeb.HFHF} gives  $\nu_3=475$ \cm\ and $\nu_6= 395$ \cm\ 
with an estimated extrapolation accuracy of 3 and 8 \cm, respectively. These
extrapolations were made using accurate values of $\nu_1+\nu_3$, $\nu_2+\nu_3$, $\nu_1+\nu_6$ and $\nu_2+\nu_6$. 
%Quack and Suhm\cite{91QuSuxx.HFHF} 
%estimated  the value of $\nu_6$ to be $376\pm 3$ $$\cm. 
Since there are
more than one estimated frequency for both $\nu_3$ and $\nu_6$  \cite{98KlWiQu.HFHF,96AnDaNeb.HFHF}, 
we present both values for comparison in %the
Table \ref{tab:obs.semi.vib}. 
%All that confirms very
%high quality of the used PES.
%Further improvement of this PES could be done
%by calculation of \ai\ points of the PES using CCSD(T) with 
%core-valence basis set and 
%inclusion of (may be) relativistic, adiabatic and higher order correlation
%corrections. Such work is at the moment under way. 

\begin{table}[]
\caption{Calculated and empirical values of inter-molecular vibrations
with $(J,K) = 0$ given in \cm. Uncertainties in units of the final digit are given in parenthesis.} 
\label{tab:obs.semi.vib}
\begin{tabular}{crr}

\hline\hline
	&	calc.	&	empirical	\\
\hline
$\Delta_{K=0}$	&	0.679	&	0.6587(1)$^a$  	\\
$\nu_4$	&	124.8	&	125.1(1)$^b$\\   
$\nu_5$	&	162.7	&	160.6(6)$^b$	\\      
$\nu_6$	&	413.0	&	395(8)$^c$	\\   
$\nu_6$	&	413.0	&	420(5)$^d$\\\
$\nu_3$	&	485.0	&	475(3)$^c$	\\          
$\nu_3$	&	485.0	&	480(10)$^d$	\\
\hline\hline
\end{tabular}
\\
$^a$ Ref. \cite{88PuQuSu.HFHF} 	\\
$^b$ Ref. \cite{96AnDaNea.HFHF}\\   
$^c$  Ref. \cite{96AnDaNeb.HFHF}\\
$^d$ Ref. \cite{98KlWiQu.HFHF}\\
\end{table}

\begin{table}[]
\caption{Calculated and experimentally determined  values of band origins with $K > 0$ in \cm.}
\label{tab:obs.Ks}
\begin{tabular}{lrr}
\hline\hline
$(v_3 v_4 v_5 v_6)$	&	calc.	&	exp. \\
\hline
(0000) $K=1$ &  36.4  &   35.4$^a$  \\
(0010) $K=1$ &  37.5  &  36.5$^a$   \\
(0000) $K=2$ & 118.0  &  116.1$^a$ \\
(0010) $K=2$ & 120.0  &  118.1$^a$ \\
(0000) $K=3$ &  235.4 &  232.6$^a$ \\
(0010) $K=3$ &  239.3 &  236.5$^a$ \\
(0010) $K=4$ &  390.5 &  386.7$^a$ \\
(0020) $K=3$ &  394.0 & 393.6$^b$  \\
(0001) $K=1$ & 399.4  & 399.8$^c$ \\
(0011) $K=1$ & 401.3  & 400.8$^c$ \\
(0001) $K=2$ & 467.2  & 465.3$^c$  \\
(0011) $K=2$ & 471.7  & 468.8$^c$  \\
(0101) $K=1$ & 507.5  & 508.1$^d$ \\
(0111) $K=1$ & 511.8  & 507.3$^d$ \\
\hline\hline
\end{tabular}
\\
$^a$ Ref. \cite{88PuQuSu.HFHF} 	\\
$^b$ Ref. \cite{90QuSuxb.HFHF} 	\\
$^c$ Ref. \cite{89PuQuSu.HFHF} 	\\
$^d$ Value obtained from energy of $\nu_6 + \nu_4 - \nu_4$, $K=1$ state \cite{87PuQuxx.HFHF} and $\nu_4$
\end{table}

%\end{longtable}

The quality of the HF dimer PES we use is
illustrated by the discrepancies between the inter-molecular vibrations
energy levels and the results of our calculations described above. All these discrepancies 
are close to the error bars in the extrapolation of the high resolution 
values of the experimental band origins. There is no way to improve 
further the PES based only on the  $J=0$ comparisons with experiment; therefore 
further developments need to use calculations that consider excited $J$s. 

Table \ref{tab:obs.Ks} presents a comparison between variationally calculated
and expe\-ri\-mentally-derived band origins of the $J$-branches of excited
$K$ levels of the ground and vibrationally excited inter-molecular modes.
The values for the ground state were taken from our 1990 study \cite{90BeKaKo.HFHF}
and the excited $K$ values of the excited vibrational states are taken
from various works by Quack and Suhm. The agreement
between calculated and experimentally derived values shown in Table \ref{tab:obs.Ks} is very good. It again confirms the high quality of the \ai\ PES used.
Even more so, since the information in %the
Table \ref{tab:obs.Ks} is directly derived
from experiment, unlike the estimated values of  
Table \ref{tab:obs.semi.vib}, the much better
agreement between calculations and experiment in Table \ref{tab:obs.Ks} emphasises 
once more the importance of high $K$ and thus excited $J$ calculations. 
The comparisons given in Table \ref{tab:obs.Ks}  could 
not be made in the original paper reporting the \ai\ PES \cite{19HuYaZh.HFHF} since that work
only performed $J=0$ calculations.

To calculate the rotational energy levels of the (HF)$_2$ dimer, we used parameters $J_{\rm max}=9$
and $K_{\rm max}=9$. We had to increase  from 10 to 500
the  number of levels written out by the program for each 
symmetry at the fourth stage, that is, when the Coriolis-decoupled $J > 0$ energy levels
are calculated. 
%The published version of WAVR4 was generalized to allow calculations with $J > 9$.
Additionally,  the angular vibrational
basis set parameter with a rotational momentum $l_{max}$ had to be 
increased from 16 to 20.
The final ro-vibrational calculations involved diagonalisation
of Hamiltonian  matrices of dimension $40\times(J+1)$. This is sufficient 
to converge the energy levels of interest to
better than 0.1 \cm. Our calculated (HF)$_2$  vibrational-rotational energy levels 
 with $J \leq 9$  are given in Table A1 of the Supplementary~ Materials.

Experimental information on the excited $J$ states of (HF)$_2$  is
usually compressed by fitting values of rotational constants (such as 
$B, D$ and $H$, see  Eq. (1) and Eq. (2)) to the observed line positions. 
This  compression is possible because the 
HF dimer behaves like a nearly linear molecule and for a certain
vibrational  and  \ka\  states, the lines of the HF dimer
are grouped in regular progressions, similar in structure to the rotational series observed for diatomic
or linear molecules. The observed rotational constants are obtained from experimental %by fitting  
line positions and calculated constants from 
 variational calculations using %by fitting
 excited $J$ energy levels. %They can also be determined from the calculation in a simplified way 
 We determined B and D constants from the fit of calculated levels with
 $J$ from 1 to 9 using Eq.~(\ref{eq:simple}).
 
% \textcolor{green}{We adopted in the present work the more reliable but a bit more approximate way} -- by deriving $B$ from the $J=1$ level and by fixing
%it and using the formula of Eq.~(\ref{eq:simple}) to derive $D$ constants from the higher
%excited $J$ variational levels, such as $J=6$ or 7. 
%\red{JT: the referees strongly disliked this procedure %which they regarded as unnecessarily approximate as doing things
%better is easy. I think we a courting further trouble if we don't change it.}
%The errors in rotational constants obtained due to such simplification are 
%less than the \textcolor{green}{potential error due to correlation of constants if both constants are fiited simultaneously to the limited number of the \ai\ calculated levels with unknown uncertainty} %errors in the \ai\ levels energies used for determination
%of the $B$ and $D$ constants. \red{JT: I am not convinced that this logic is really a justification of the above approximation.} 
%\red{The JCP referee really did not like the use of only J=1 to fix B.
%We need to either justify it or do a better fit which uses all J's to %determine B and D (should be easy enough).}
%$H$ constants can be  obtained approximately from further, higher $J$ calculations in the same manner. 

Table \ref{tab:obs.calc.const} presents  values for our computed $B$
and $D$ constants
and, where possible,  compares them with empirical values. We will 
refer to these collectively as 'spectroscopic constants' here and elsewhere.
For convenience 
and completeness Table \ref{tab:obs.calc.const} also presents
 experimental values for $K=4$ of the ground state and some excited inter-molecular 
vibrational states  obtained in %the 
Section \ref{sec:analysis}
of this paper;
other experimental values %for $v=0$, $B$ and $D$  and $\Delta_K$ constants
are taken from the literature, mostly from the papers by Quack and Suhm.
%More details of these values and specific references are given in the  section \ref{sec:analysis}.

%\begingroup
%\renewcommand*{\arraystretch}{1.5} 

\begin{longtable}{ccccccl}
	\caption{Calculated and experimental rotational constants (powers of ten are given
	in parenthesis).}
%	When $D$ 
%	values are given in \cm, powers of ten are given in parenthesis}
	\label{tab:obs.calc.const} \\
%	\hline
%	\multicolumn{3}{c}{Уровень}	&	Obs.	&	Calc.	\\
%	\hline\endfirsthead
	\hline\hline
		State	&	Units	& $B_{calc}$ & $B_{exp}$ &	$D_{calc}$ &	$D_{exp} $	& Exp. source\\
	\hline\endfirsthead
	\hline\\[3pt]
		State	&	Units	& $B_{calc}$ & $B_{exp}$ &	$D_{calc}$ &	$D_{exp}$ 	& Exp. source\\
	\hline\endhead
	\hline
%	\multicolumn{6}{c}{\textit{Continued on next column...}} \\
	\endfoot
	\endlastfoot
	(0000)	&		&		&		&		&	&   \\
$K=0$ &\cm&	0.2154	&	0.2167   &	2.26($-6$)	&	2.06($-6$)	&  Ref. [\citenum{88PuQuSu.HFHF}] \\[1.75pt]
		    &	MHz	    &	6457.5	&	6496.9   &	0.0633	    &	0.0618		&   \\[1.75pt]
$K=1$&\cm&   0.2158	&	0.2179   &	1.92($-6$)	&	1.99($-6$)	&  Ref. [\citenum{88PuQuSu.HFHF}] \\[1.75pt]
		    &	MHz	    &	6469.5	&	6531.3 	 &	0.0577	    &	0.0596		&   \\[1.75pt]
$K=2$&\cm&	0.2174	&	0.2186 	 &	1.89($-6$)	&	1.94($-6$)	&  Ref. [\citenum{88PuQuSu.HFHF}] \\[1.75pt]
		    &	MHz 	&	6517.4	&	6553.7   &	0.0567	    &	0.0582	&   	\\[1.75pt]
$K=3$&\cm&	0.2169	&	0.2181   &	1.47($-6$)	&	1.76($-6$)	&  Ref. [\citenum{88PuQuSu.HFHF}]  \\[1.75pt]
		    &	MHz	    &	6502.4	&	6538.2 	 &	0.0441	    &	0.0529		&   \\[1.75pt]
$K=4$&\cm&	0.2150	&	0.2175   &  6.14($-5$)&	-6.01($-5$)	&  This work \\[1.75pt]
		    &	MHz	    &	6444.6	&   6498.6   & 	1.84	    &  	-1.8		&   \\[1.75pt] 
\hline \\[3pt]
	(0010)	&		&		&		&		&		&   \\ [1.75pt]
$K=0$&\cm&	0.2153 	&	0.2166 	&	2.41($-6$)	&	2.04($-6$)	&   Ref. [\citenum{88PuQuSu.HFHF}]  \\[1.75pt]
		    &	MHz	    &	6451.5 	&   6492.8  &	0.0625	    &	0.0612		&   \\[1.75pt]
$K=1$&\cm&   0.2177	&   0.2177 	&	2.08($-6$)	&	1.97($-6$)	&  Ref. [\citenum{88PuQuSu.HFHF}]  \\[1.75pt]
		    &	MHz	    &	6525.5	&	6526.8 	&	0.0625	    &	0.0591		&   \\[1.75pt]
$K=2$&\cm&	0.2173	&	0.2185 	&	1.86($-6$)	&	1.95($-6$)	&   Ref. [\citenum{88PuQuSu.HFHF}]  \\[1.75pt]
		    &	MHz	    &	6514.3	&	6551.1  &	0.0558	    &	0.0586		&   \\[1.75pt]
$K=3$&\cm&	0.2176	&	0.2188 	&	1.60($-6$)	&	1.87($-6$)	&  Ref. [\citenum{88PuQuSu.HFHF}]  \\[1.75pt]
		    &	MHz	    &	6522.7	&	6558.9	&	0.0479	    &	0.0561		&   \\[1.75pt]
$K=4$&\cm&	0.2181 	&	0.2194  &  1.74($-6$)       &	2.28($-6$)	&  Ref. [\citenum{88PuQuSu.HFHF}]  \\[1.75pt]
		    &	MHz	    &	6539.1	&   6576.9  & 	0.052	&	0.068  	&   \\	[1.75pt]
\hline \\[3pt]
	(0100)	&		&		&		&		&		&   \\	[1.75pt]
	$K=0$   &	\cm	    &	0.2119	&	0.2140	&	3.04($-6$)&	4.34($-6$)& This work   \\	[1.75pt]
	        &   MHz	    &	6353.0	&	6415.5	&	0.0910	&	0.130		&   \\	[1.75pt]
%	$K=1$	&	\cm	    &	0.2112	&	        &   2.80(-6)&		&   \\
%		    &	MHz	    &	6330.1	&           &   0.0840	&		&   \\
\hline \\ [3pt]
	(0110)	&		&		&		&		&		&   \\	[1.75pt]
	$K=0$ 	&	\cm	    &	0.2104	&	0.2161	&	2.71($-6$) &	5.07($-6$)& This work  	\\	[1.75pt]
		    &	MHz	    &	6306.3	&	6478.5	&	0.0813	 &	0.152		&   \\	[1.75pt]
%	$K=1$	&	\cm	    &	0.2099	&	        &  2.21(-6)  &		    	&   \\
%		    &	MHz	    &	6293.8	&	        &  0.0663    &				&   \\
\hline\\ 	[3pt]
	(1000)	&		&		    &		    &		    &		&   \\	[1.75pt]
	$K=0$   & \cm   &	0.2052	& 0.2129    & 3.24($-6$)	&1.88($-4$)		& This work  \\	[1.75pt]
		    &	MHz	&	6350.2	&	6383.9  & 0.0971	&  5.63	&   \\	[1.75pt]
\hline \\	[3pt]
	(1010)	&		&		&		&		&	&   	\\	[1.75pt]
	$K=0$	&	\cm	&	0.2099	& 0.2002 &  2.97($-5$) &1.52($-4$)	&  This work 	\\	[1.75pt]
		    &	MHz	&	5922.3	& 6000.4 &  0.890	& 4.57	&   \\	[1.75pt]
\hline \\	[3pt]
	(0001)	&       &		    &		    &		        & 	&   \\	[1.75pt]
	$K=0$	&	\cm	    &	0.2154	&	        & 1.02($-5$)  & 	&   \\	[1.75pt]
	        & MHz	    &	6460.1	&	        & 0.306	    & 	&   \\	[1.75pt]
$K=1$&\cm&	0.2132 	&0.2105  &1.13($-5$) & 2.5($-6$) 	&  Ref. [\citenum{89PuQuSu.HFHF}] \\	[1.75pt]
	        &  MHz	    &   6393.7  &	6310.6  &	0.34	 & 0.07   	&   \\	[1.75pt]
$K=2$&\cm&	0.2127	&0.2113	&1.03($-5$) & 2.1($-6$) 	&  Ref. [\citenum{89PuQuSu.HFHF}] \\	[1.75pt]
	        &   MHz	    &	6375.4	&	6334.6	&   0.31     & 0.06    	&   \\	[1.75pt]
%	$K=3$	&	\cm	    &		    &	    	&			\\
%	        &  MHz   	&	    	&		    &			\\
\hline \\	[3pt]
(0011)	&       &	&   		&   		&          	&   \\	[1.75pt]
$K=0$ &	\cm	    &	0.2157	&	        & 3.09($-6$)  &  			&   \\	[1.75pt]
	    &  MHz	    &	6466.9	&	        & 0.0926	&		 	&   \\	[1.75pt]
$K=1$&\cm&	0.2132	&0.2103  & 1.08($-5$)	& 2.5($-6$)  	&  Ref. [\citenum{89PuQuSu.HFHF}] \\	[1.75pt]
	    &   MHz	    &	6385.9	&	6304.6  & 0.324	    & 0.07    	&   \\	[1.75pt]
$K=2$&\cm&	0.2140	&0.2129	& 0.99($-6$)    & 2.05($-6$)&  Ref. [\citenum{89PuQuSu.HFHF}] \\	[1.75pt]
	    &  MHz	    &	6415.9	&	6382.5	& 0.30    & 0.0615		&   \\	[1.75pt]
\hline\\	[3pt]
(0101)	&           &	&		    &	        	& 	&   \\	[1.75pt]
%$K=0$	&	\cm	    &	0.2080	&    	    &	    &		&   \\
%	    &   MHz	    &	6236.5	&	        & 0.306	&		&   \\
$K=1$	&	\cm	    &	0.2080	&  0.2133   & 7.67($-6$)& 3.74($-6$) & $^a$ \\	[1.75pt]
	    &   MHz	    &	6236.5	&  6334.6	& 0.23  & 0.112		&   \\	    	[1.75pt]
\hline \\	[3pt]
(0111)	&   &	&		    &		    &   	&   \\	[1.75pt]
%	$K=0$ &	\cm	    &	0.2072	&  	        &           & 1.02(-5)  	&   \\
%	    &   MHz	    &	6211.3	&           & 0.306	    &		&   \\
	$K=1$ &	\cm	    &	0.2072	&  0.2114   & 9.54($-6$) & 5.03($-6$) & $^a$  \\	[1.75pt]
	    &   MHz	    &	6211.3	&  6337.6   & 0.286	    & 0.151		&   \\	    	[1.75pt]
\hline \\	[3pt]
(0020)	    &      &	&		    &		    &  	&   \\	[1.75pt]
$K=3$&\cm& 0.2227   &  0.2217   &  4.5($-6$)  & 1.67($-6$)	&  Ref. [\citenum{90QuSuxb.HFHF}] \\	[1.75pt]
	        & MHz	    &  6679.4	&  6646.3   &  0.13	    & 0.050		&   \\	[1.75pt]
\hline\hline
\multicolumn{7}{l}{$^a$Constants calculated from $\nu_4$ constants determined }\\
%\textcolor{blue}{(remove Fig.1 from the table)}  
\multicolumn{7}{l}{in this work and values for $\Delta{X}$ ($X=B,D$) in Ref. [\citenum{89PuQuSu.HFHF}]} \\
\end{longtable}

%\endgroup

Table \ref{tab:obs.calc.const} shows that the ground state $B$ constants
taken from the Belov \etal\ \cite{90BeKaKo.HFHF}
and calculated variationally in the present paper coincide to within 
about 30 -- 50 MHz. 
This agreement between $B_{obs}$ and $B_{calc}$ is unexpectedly good.

The impressive accuracy of prediction of $B$ and $D$ constants associated with the
tunneling rotation spectrum of the ground vibrational state of the HF dimer %, presented in Ref. \cite{90BeKaKo.HFHF} given in %the Table \ref{tab:obs.calc.const},
and also of the $B$ values of some excited vibrational states 
taken from the papers by Quack and Suhm, paves the way for us to
apply the results of these calculations to an analysis of unassigned
lines in the HF dimer tunneling-rotation spectra. These lines belong to 
excited inter-molecular vibrations and higher $K$ states  of the ground 
vibrational state; for instance, 
the high resolution spectra of tunneling-rotational lines in $v_4 =1$ and $v_3=1$ %MT: or $\nu_4$ and $\nu_3$ 
vibrational states 
are assigned for the first time here (Section 4). 
For other vibrational states,
high resolution spectra have been analysed by Quack and Suhm \cite{87PuQuxx.HFHF,89PuQuSu.HFHF,90QuSuxb.HFHF,90QuSuxx.HFHF}.
%of these states is obtained here for the first time.
%all four inter-molecular vibrations, i.e. $v_4=1$, $v_5=1$, $v_3=1$ 
%and $v_6=1$.

%\textbf{(Vladimir's part)}
\begin{figure}[ht!]
	\includegraphics[scale=0.65]{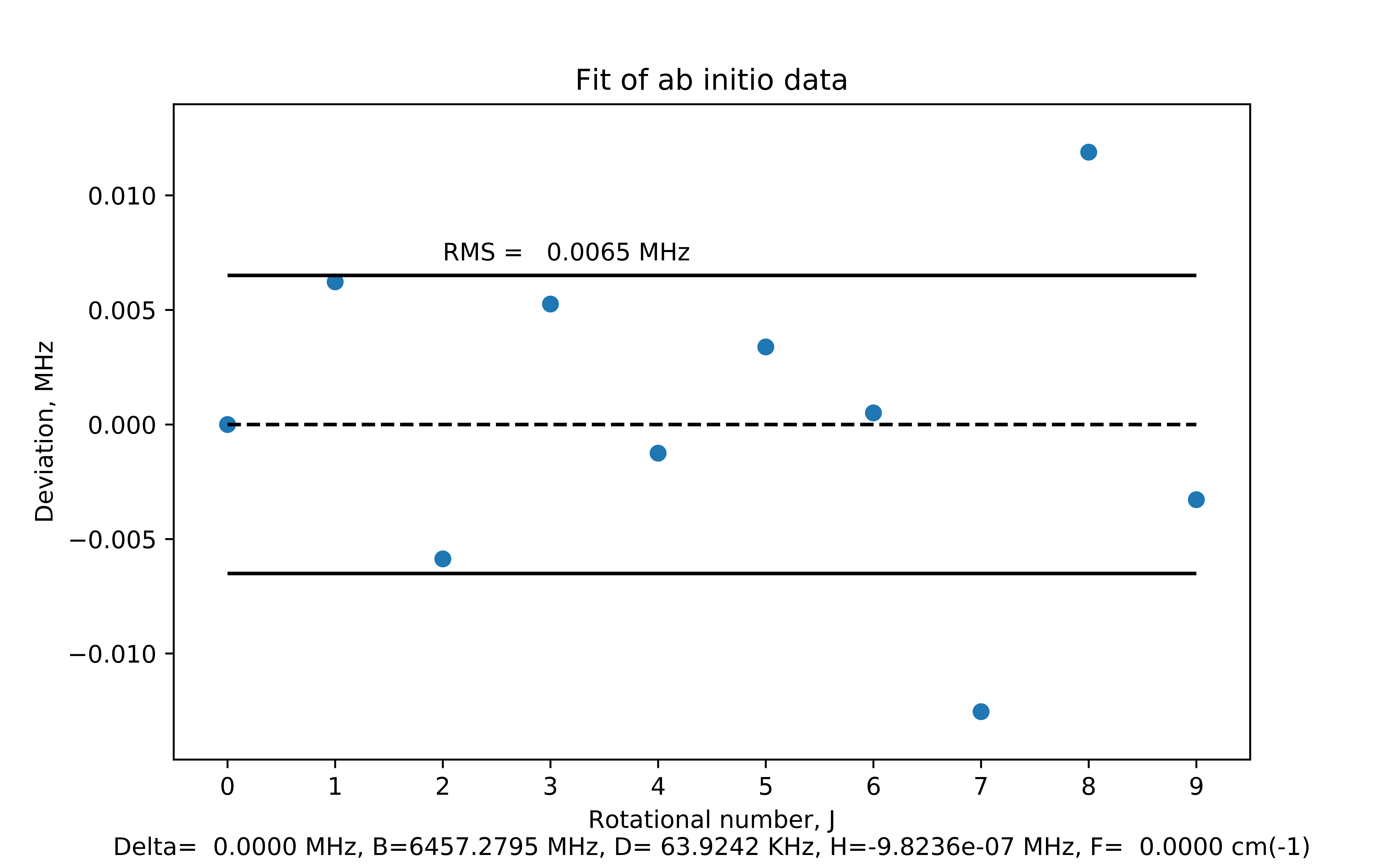}
	\caption{Difference between the \ai\ energies and ones obtained 
	from the fit for $(v_3 v_4 v_5 v_6)=(0~0~0~0)$ band of (HF)$_2$, $K=0$ 
	%\textcolor{blue}{(see my note about the figure in that PDF)}
	}
	\label{fig:0000-k0}
\end{figure}

\begin{figure}[ht!]
	\includegraphics[scale=0.65]{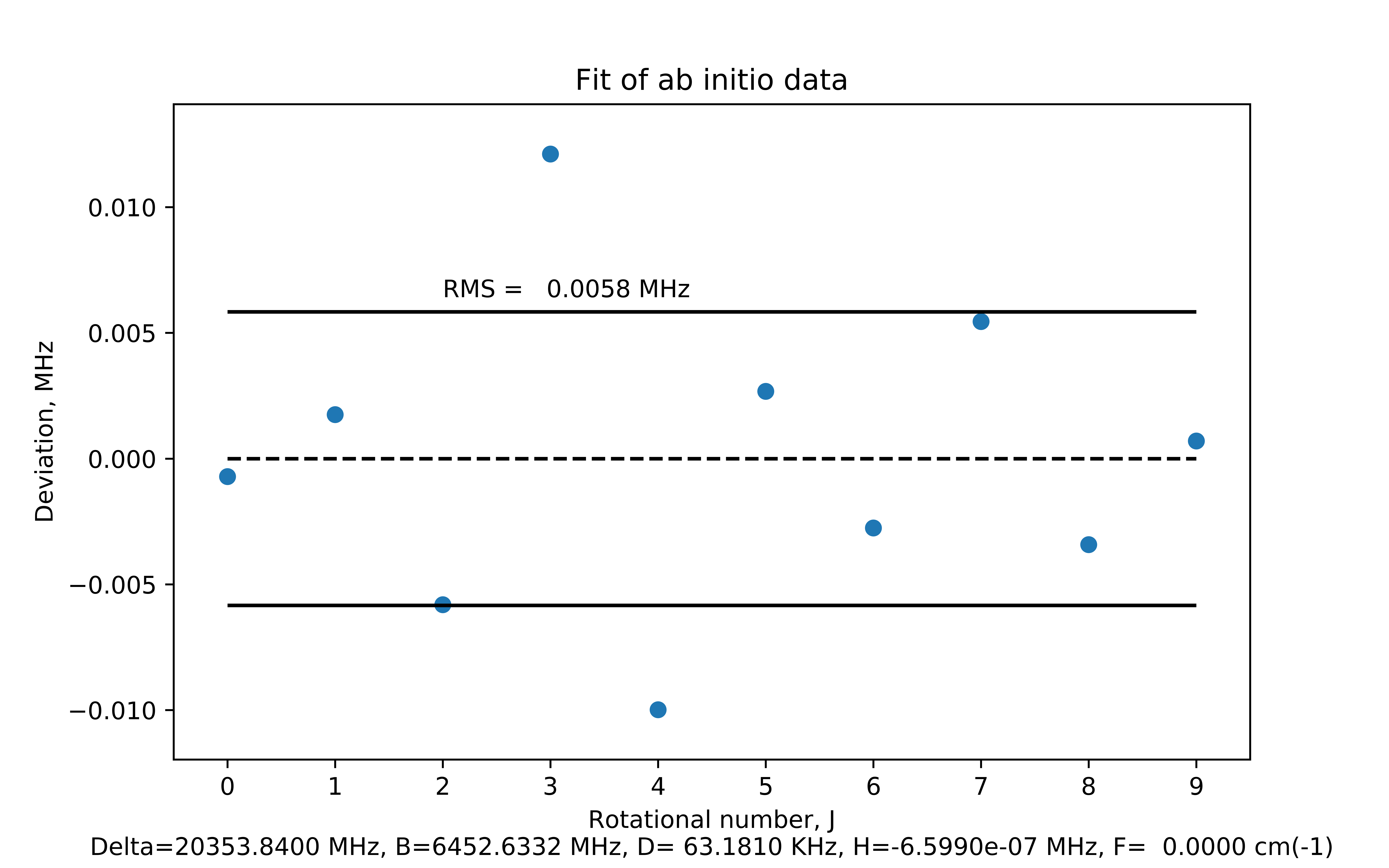}
	\caption{Difference between the \ai\ energies and ones obtained 
	from the fit for $(v_3 v_4 v_5 v_6)=(0~0~1~0)$ band of (HF)$_2$, $K=0$ 
	%\textcolor{blue}{(see my note about the figure in that PDF)}
	}
	\label{fig:0010-k0}
\end{figure}

We also performed  a least-squares fit to the \ai\  energies using the 
formula of Belov \etal ~\cite{90BeKaKo.HFHF}
\begin{equation}
\label{eq:BHD_full}
    \begin{split}
	    E(J, K, \nu) = \Delta_{K}\delta_{\nu,1} + F_K + B_J^{\nu}J(J+1) \\
	    - D_J^{\nu}J^2(J+1)^2 + H_J^{\nu}J^3(J+1)^3 .
	\end{split}
\end{equation}
Results of this fit are presented in Figures \ref{fig:0000-k0} and \ref{fig:0010-k0}.
Small residuals (less than 100 kHz) confirm that the formula \ref{eq:simple} 
used for the fit is appropriate and 
we can thus safely use the results of this fit to represent the 
energies obtained by our  \ai\  calculations. In other words
the values of $B$ and $D$ given in  
Table \ref{tab:obs.calc.const} give an excellent representation of 
the values calculated when used with Eq.~(\ref{eq:BHD_full}) and effect 
on energy levels related to H-constants may be omitted.

\section{Experimental details of the observation of sub-millimeter spectrum
	of HF dimer in the gas phase}
\label{sec:experiment}

 Thirty years ago some of us published a sub-millimeter study \cite{90BeKaKo.HFHF} of the (HF)$_2$  spectrum. At the time it was not possible to completely assign this spectrum.
 Before discussing new assignments we have been able to make using 
our variational calculations (Section \ref{sec:analysis}), we first give some  experimental details.

The broad-band (180-380 GHz) spectrum of the HF dimer was recorded %by us 
at the end 
of eighties using a spectrometer with a backward wave oscillator 
(BWO)  as a radiation source \cite{Istok-site}
%\footnote{\href{http://istokmw.ru/}{istokmw.ru}} 
and Radio-Acoustic Detection of absorption (RAD), 
which we call the RAD-3 spectrometer. A detailed description of the instrument 
can be found on the Laboratory website\footnote{\href{https://mwl.ipfran.ru/}{http://mwl.ipfran.ru}}
and in our report on a similar study of the H$_2$O-HF
dimer \cite{07BeDeZo.H2O-HF}. 
A brief description of the experiment considered was given in the corresponding section 
of Ref. \cite{90BeKaKo.HFHF}. For the reader's convenience we 
repeat here the most important 
details with some other complementary information which was used in the
present analysis of the spectrum.

Equilibrium gas-phase HF dimer spectra were observed at temperatures of 
about 200 - 220 K in a static pressure stainless
steel cell at pressures between 0.5 – 1.5 Torr. HF vapor was 
obtained by thermal decomposition of KFHF. The intensity of the observed 
spectrum depended on pressure and temperature. The prominent spectral 
lines appeared and grew up only with cooling the cell down and their
intensity also increased with increased pressure  as  expected 
from the general pressure and temperature dependence characterized by the dimerisation constant. %MT: The constant itself does not depend on pressure...
The optimal temperature was found to be near 210 K. Neither the HF
gas generator nor the cell thermostat allowed active control 
of experimental conditions. Because of this the spectral intensity
in the repeated recordings varied from day to day but relative 
intensity of lines within one recording was preserved. 

\begin{figure*}[ht]
	\includegraphics[scale=0.22]{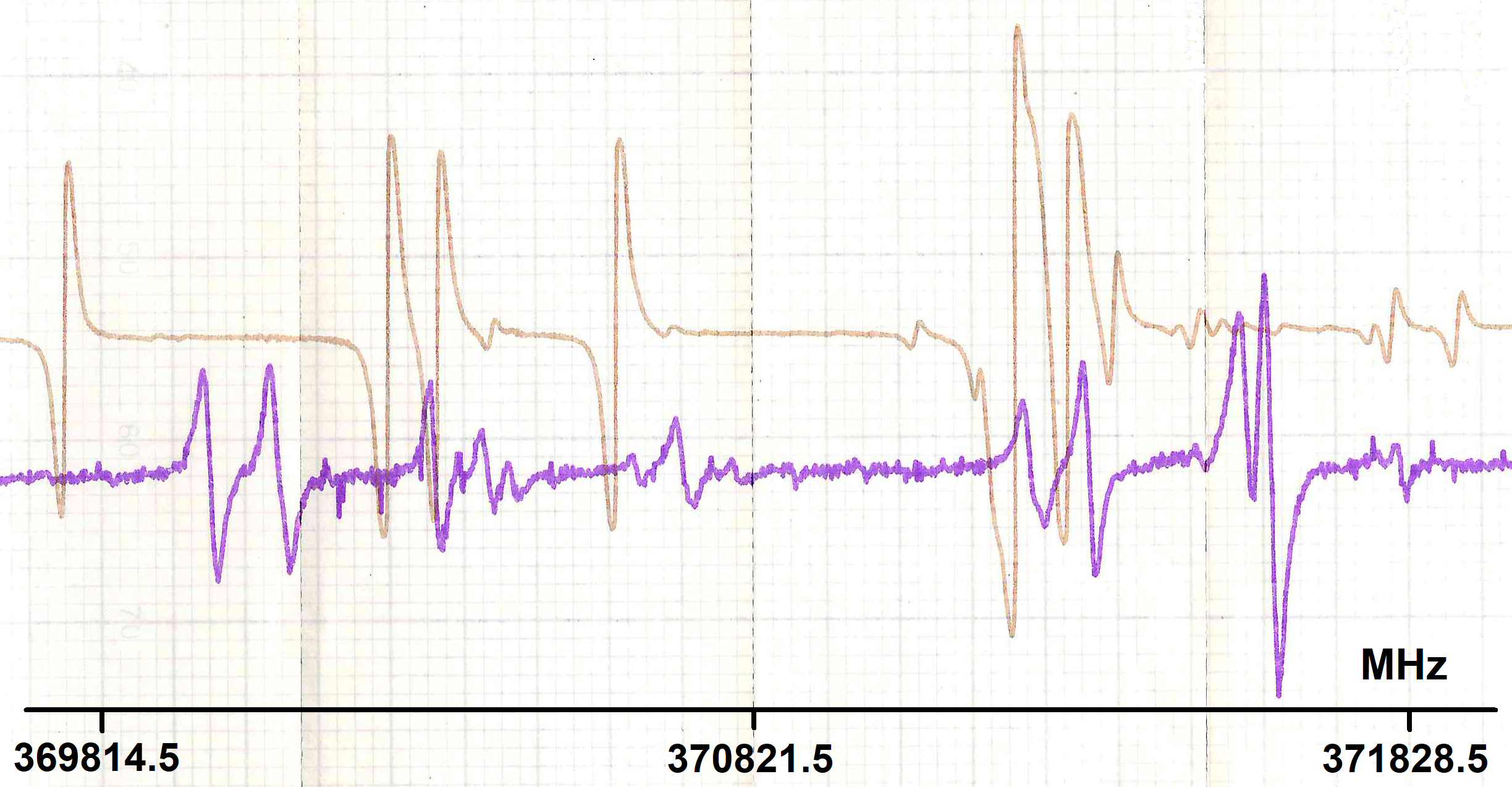}
	\caption{Small portion  of recorded in the late 80-ties 
	tunneling-rotation spectra of (HF)$_2$ in violet (lower spectrum) and calibration SO$_2$ in light brown (upper spectrum) 
	%\textcolor{blue}{(see my note about this figure in that PDF)}
	}
	\label{fig:exp-expl}
\end{figure*}

The spectrum was recorded 
continuously (without gaps) 
covering the  range from 180 up to 385 GHz. 
The continuity is crucial 
for unambiguous assignment of observed lines. At the time 
of the experiment,
such broadband continuous recordings were not possible 
in this spectral range as the 
radiation frequency was controlled using a phase locking loop systems with poor scanning ability  %was fixed using a frequency standard 
(see review  \cite{12KrTrBe.methods} 
and references therein). Frequency measurements in 
the spectrometer were calibrated using 
the lines from a SO$_2$ reference spectrum \cite{98BeTrKo.SO2}.
The mean square value of the frequency uncertainty  was determined 
in preliminary tests as 0.5 MHz for strong isolated monomer lines. 
The uncertainty 
was somewhat larger in the case of weak and pressure broadened 
HF dimer lines. 
However, we consider a value of a few MHz as a reasonable estimate 
of the possible error 
in the frequency measurements for most lines. 
Figure \ref{fig:exp-expl} 
%(Коля выбери красивый кусок, но чтобы он не повторял 
%опубликованные в РФ и JMS) 
presents a portion of the observed HF dimer spectrum  recorded  simultaneously with a SO$_2$ 
reference spectrum. 
%Measured frequencies of the 
%(HF)$_2$ spectrum and tabulated frequencies of the reference 
%SO$_2$ spectrum are shown next to the lines. 
Unfortunately, only half of the recorded spectrum has survived 
the transition from large specialized counting machines 
to personal computers. Only recordings in the range 
of 268 -- 385 %Коля, проверь.
GHz are now available. Analysis of this spectrum is 
given in the next section.

It is our view that repeating the experiment 
using a modern, more 
sensitive and more accurate apparatus alongside 
extension of the spectral range to higher frequencies, where the
characteristic Q-branches are located, is  now
important  for obtaining  data for comparison with 
 high accuracy \ai\ 
calculations. However, this would have to  be the subject of a specialized project
with adequate funding. 

\section{Analysis of the sub-millimeter spectrum}
\label{sec:analysis}

Accurate \ai\ prediction of highly
rotationally excited $J$ and $K$ energy levels are essential
for the analysis of the sub-millimeter tunneling-rotation spectrum 
of the HF dimer in the 
limited region between 268 and 385 GHz.  Quack and Suhm 
 were able to assign the lines of the far IR HF dimer
spectrum, \cite{87PuQuxx.HFHF,89PuQuSu.HFHF,91QuSuxx.HFHF} 
since the IR 
experimental spectra with much wider coverage provided information
on all the P, Q and R branches. Moreover, ro-vibrational lines belonging to the different
fundamental bands of the different inter-molecular modes, $\nu_3$, $\nu_4$,
$\nu_5$ and $\nu_6$, occur in distinct, non-overlapping regions
of the spectrum. Let us briefly summarise their assignments. %of the exited inter-molecular branches. 
Von Puttkammer and Quack \cite{87PuQuxx.HFHF} studied
%putkamQ, Von Putkamer K, Quack M, Mol.Phys.,vol62,1047 (1987)
the region between 350 and 550 \cm, where the observation of two Q-branches
helped them to assign the $\nu_6$, $K=1 \leftarrow 0$ band with a Q-branch at around 400 \cm.
In the same paper they observed another clearly resolved $P,\ Q,\ R$ structure
with a Q-branch at 383 \cm. In a subsequent paper \cite{91QuSuxx.HFHF},
Quack and Suhm attributed this structure to the  $\nu_6+\nu_4 -\nu_4$ hot band.
Von Puttkammer \etal~\cite{89PuQuSu.HFHF} 
analyzed  the $\nu_6$, $K=2 \leftarrow 1$ subband at 465 \cm, and Quack and Suhm \cite{90QuSuxb.HFHF}
investigated $K=2$ and 3 subbands of the $\nu_5$ band and 
predicted the 6.83 \cm\ 
sub-millimeter rotation-vibration tunneling band between excited 
vibrational states.
Figure 4 of their paper \cite{90QuSuxb.HFHF} is a 
diagram of all known and estimated
high $K$ energy levels of $\nu_4$, $\nu_5$ and $\nu_6$ bands of the HF dimer. 
This figure facilitates 
understanding of the energy levels
presented in Tables \ref{tab:obs.semi.vib} and \ref{tab:obs.Ks}. 

To summarize the previously published work,
mostly by Quack, Suhm and co-authors, experimental 
information on inter-molecular vibrations with $J=0$ and excited $K$ the 
following is known (Tables \ref{tab:obs.semi.vib} and \ref{tab:obs.Ks}). $K=0$ is unknown 
from analysis of high resolution spectra  for all 
inter-molecular states. However, different 
extrapolation procedures  have allowed frequencies of all four low-lying vibrations to be estimated with
differing levels of accuracy (Table \ref{tab:obs.semi.vib}). In particular, for  
the $\nu_4$ band  indirect experimental information on the 125 \cm\ 
band origin, derived from $\nu_1+\nu_4$ and $\nu_2+\nu_4$ data 
extrapolation
by Anderson \etal\ \cite{96AnDaNea.HFHF}, and from  an analysis of the $J \geq 30$ 
resonance between  $K=2$ of 
the ground state and $K=0$ of the $v_4 =1$ state by Quack and Suhm 
\cite{91QuSuxx.HFHF}.

Comprehensive information
on the $K \leftarrow K-1$ bands for $K=1$ and 2 was obtained by Quack and Suhm 
for $\nu_5$ \cite{90QuSuxb.HFHF}
and $\nu_6$ \cite{89PuQuSu.HFHF}
states. %bands. 
However, no high resolution information on $\nu_3$
and $\nu_4$ has been published. In what follows we present the first high
resolution experimental assignment of lines belonging to  these two states.%bands. 

\begin{table}[h]
\caption{Observed frequencies (in MHz) for the $K=0$ branch of the tunneling-rotation spectrum of HF dimer
in the $\nu_4$ excited vibrational state.}
\label{J_J}
\begin{tabular}{cccrc}
\hline\hline
Frequency    &   $J$'  & $J$" & obs.-fit. \\ 
\hline
274791.3  &    15  & 14 & -0.1 \\
285878.5  &    16  & 15 & -0.5 \\ 
296872.7  &    17  & 16 &  1.5 \\
307772.9  &    18  & 17 & -0.7 \\ 
318591.5  &    19  & 18 &  0.1 \\
329329.8  &    20  & 19 & -1.0 \\
339995.3  &    21  & 20 & -0.1 \\
350590.3  &    22  & 21 &  0.4 \\
361118.4  &    23  & 22 &  1.8 \\
371574.7  &    24  & 23 & -2.1 \\
381970.6  &    25  & 24 &  0.6 \\
\hline\hline
\end{tabular}
\end{table}

\begin{table}[!h]
\caption{Values (in MHz) of the effective rotational $B$ and $D$ constants 
as well as the tunneling splitting constant $\Delta_K$ (Eq.(\ref{eq:BHD_full})) 
of the $K=0$ tunneling-rotation spectrum of $\nu_4$.}
\label{BDH}
\begin{tabular}{lll}
\hline
\hline
    & Calc. &  Obs. \\
\hline
$\Delta_0$  & 88659    &   94494 \\
\hline
\multicolumn{3}{c}{Upper} \\
\hline
$B$           &  6306.3  &     6433.9  \\
$D$            &  0.0813&       0.167   \\
\hline
\multicolumn{3}{c}{Lower} \\
\hline
$B$           &  6353.0  &    6488.8   \\
$D$           &    0.091 &  0.172   \\
\hline
\hline
\end{tabular}
\end{table}

\begin{table}[ht]
\caption{$K=4$ branch of tunneling-rotation spectrum 
in the ground vibrational state in MHz.}
\label{K=4J_J}
\begin{tabular}{cccr}
\hline\hline
Frequency    &    $J$'  & $J$" & obs.-fit\\
\hline
280614.3  &     5  &  4 & -0.1\\
294326.6  &     6  &  5 &  1.2\\ 
308144.8  &     7  &  6 & -3.4\\
322151.6  &     8  &  7 &  4.9\\ 
336419.3  &     9  &  8 & -3.7\\
351119.4  &    10  &  9 &  1.1\\
366409.2  &    11  & 10 &  0.2\\
382504.2  &    12  & 11 &  0.1\\
\hline\hline
\end{tabular}
\end{table}

\begin{table}[!ht]
\caption{Values (in MHz) of effective rotational $B$ and $D$ constants 
as well as the tunneling splitting constant $\Delta_K$ (Eq.(\ref{eq:BHD_full})) 
of the $K=4$ tunneling-rotation spectrum of HF dimer in the ground
vibrational state.}
\label{K=4BDHnu0}
\begin{tabular}{lll}
\hline
\hline
    & Calc. &  Obs. \\
\hline
$\Delta_4$  & 225051    &   213545 \\
\hline
\multicolumn{3}{c}{Upper} \\
\hline
$B$           &  6539.1  &     6576.9  \\
$D$           &  0.052&      0.068   \\
\hline
\multicolumn{3}{c}{Lower} \\
\hline
$B$           &  6444.6  &    6498.6   \\
$D$           &    1.84 &   -1.8   \\
\hline
\hline
\end{tabular}
\end{table}

\begin{table}[!ht]
\caption{$K=0$ branch of tunneling-rotation spectrum of HF dimer in the $\nu_3$ state (in MHz) }
\label{nu3J_J}
\begin{tabular}{cccr}
\hline\hline
Frequency       & $J$'   & $J$'' & obs.-fit. \\
\hline
325666.6    &	 1 & 0	&  1.5 \\
336768.3    &	 2 & 1	& -6.3 \\
346933.0    &	 3 & 2	&  4.3 \\
356077.7    &	 4 & 3	&  8.1 \\
364146.9    &	 5 & 4	& -14.2 \\
371203.5    &	 6 & 5	&   5.7 \\
377198.7    &	 7 & 6	&   0.8 \\
382203.6    &	 8 & 7	&  -0.9 \\
\hline\hline
\end{tabular}
\end{table}

\begin{table}[!ht]
\caption{Values (in MHz) of effective rotational $B$ and $D$ constants 
as well as the tunneling splitting constant $\Delta_K$ (Eq.(\ref{eq:BHD_full})) 
of the $K=0$ tunneling-rotation spectrum in the $\nu_3$ state.}
\label{K=4BDHnu3}
\begin{tabular}{lll}
\hline
\hline
    & Calc. &  Obs. \\
\hline
$\Delta_0$  &317824&    313681  \\
\hline
\multicolumn{3}{c}{Upper} \\
\hline
$B$           &  5957.4  &     6000.5  \\
$D$            &   0.89 &       4.57    \\
\hline
\multicolumn{3}{c}{Lower} \\
\hline
$B$           &  6232.7  &    6383.91   \\
$D$           &    0.097 &  5.63   \\
\hline
\hline
\end{tabular}
\end{table}

As an example of the use of calculated HF dimer %inversion 
tunneling splittings, band origins and rotational
constants, we assigned part of  the $K=0$ R-branch %inverse
tunneling-rotational 
transitions within the exited fundamental inter-molecular vibration $\nu_4$. 
The $K=0$, $\nu_4$ state lies at about 125 \cm, which is close to the  ground  
 state  $K=1$ (35 \cm) and $K=2$ (116 \cm) states.
 So the Boltzmann factor suggests it should be the strongest of
the remaining unassigned bands. Manually we have  a set of 11 
lines covering all the available  experimental 
region of 268--385 GHz. Using seven constants ($B$, $D$ and $H$ rotational constants 
for upper and lower tunneling  split states and the value of 
this splitting) as fitting parameters, the lines are described 
with a standard deviation of
1.7 MHz corresponding to an estimated experimental accuracy of 2-5 MHz. 
If we omit the two $H$ constants, the standard deviation increases 
to 4.5 MHz which is still inside experimental accuracy.
As the number of fitted lines is not much larger than the number
of constants when H constants are included, the $H$ constants are poorly determined. 
For this reason only fit with $H$ constants fixed to zero are presented

The fit was made assuming that the inversion splitting of the $K=0$,
$\nu_4$ state is close to the calculated value of about 3 \cm\ (90 GHz). 
This value, plus the
values of the $B$ rotation constants, determine the upper and lower 
$J$'s of the measured transitions as in Eq. (2).
Changing the splitting by 12 GHz leads to a change in  the $J$'s by one. 
The correct $J$ values  for the observed transitions could not be 
determined from available experimental transitions in the 
region 268--385 GHz. Fits to the experimental frequencies with 
$J$'s differing by one give practically the same standard deviation,
and similar
values for the rotation constants. 
To empirically determine the correct $J$ values  one has to observe either the Q-branch or 
the beginning of the R-branch in the centimeter wavelength region.
Fit residuals for the tunneling-rotation $K=0$ branch lines in the $v_4=1$ state are
presented in Table \ref{J_J}. A comparison of experimentally derived 
$B$ and $D$ constants with those obtained variationally is given in
Table \ref{BDH}. Table \ref{tab:obs.calc.const} presents a corresponding comparison 
for all known $v$ and $K$ states. The residuals of \ai\ calculations for the observed frequencies
are of order  0.1 \cm\ (3000 MHz).

This first determination of $B$ and $D$ constants for the $\nu_4$ state helped us
to obtain another important piece of experimental information. In their study
on the first IR hot spectrum of combination band $\nu_6+\nu_4 - \nu_4$
Von Puttkamer and Quack \cite{87PuQuxx.HFHF} 
%\red{Tony comment: Is this correct? I could not find any %mention of $\nu_6+\nu_4 - \nu_4$
%combination band in the paper by Von Puttkamer, 
%Quack and Suhm.}
could determine only $\Delta B$
and $\Delta D$ constants as they did not of a 
know the $B$ and $D$ constants of the lower state $\nu_4$, so only differences
between upper and lower state constants could be determined. Our determination
of  the $\nu_4$ $B$ and $D$ constants allows us to determine the $B$ and  $D$ 
constants of the upper combination state of $\nu_6+\nu_4$. These newly
determined values are used in  Table \ref{tab:obs.calc.const} for comparison with the 
variationally determined $B$ and $D$ constants.

%Let us now consider the $v_3=1$ tunneling-rotation band assignment.
Let us now consider the assignment of tunneling-rotation lines in the $\nu_3$ state. 
This mode %band 
is the highest energy inter-molecular vibration (about 480 \cm ) 
and its lines  should be among the weakest in the recorded spectrum. 
The inversion splitting of the $v_3=1$ state %band 
is calculated to be about 318 GHz.% or in the first half of BWO-30 region. 
We were able to find a series of lines starting at 325 GHz. Fits to this series of 8 lines with 5 constants ($B$ and $D$ constants for upper and lower states and the  tunneling splitting) gives a standard deviation of 6.7 MHz. Observed values of the
frequencies and the residuals of the fit are given in %the
Table 8 and constants obtained from the fit and our
variational calculations are given in Table 9.

 The last series of lines %branch 
 which we assign in the 268 - 385 GHz 
 tunneling-rotation spectrum was the $K=4$ branch %band 
 of the ground state. Table \ref{K=4BDHnu0} gives lines 
 of the branch %R-band 
 for $J$ ranging
 from 4 to 11  together with the residuals between
 observed and fitted frequencies using  the $B$, $D$ and $\Delta_K$ model. $B$ 
 and $D$ constants of the
 upper state given in Table \ref{K=4BDHnu0} were fixed to the values reported by 
 %due 
 to Quack and Suhm~\cite{88PuQuSu.HFHF}, obtained from the $K=4 \leftarrow 3$ band. $\Delta_K$,
  $B$ and $D$ constants are determined in the fit of Eq.(2) to the data of Table \ref{K=4J_J}.
 The standard deviation of this fit is 4.7 MHz. 

 To summarize  our new assignments of the tunneling rotation sub-millimeter 
 spectrum: lines belonging to previously unobserved %previously unknown 
 $\nu_3$ and $\nu_4$ %bands
 states are found  %observed 
 and assigned for their $K=0$
 branches. The $K=4$ branch of the ground vibrational state had only a single known tunneling state 
 \cite{89PuQuSu.HFHF} and now both tunneling states involved in
 the tunneling rotation band have been observed, assigned and $B$, $D$ and
 $\Delta_K$ constants determined. The variationally calculated $B$ and $D$ constants
 together with the constants obtained from the fit are given in the tables
 of Section \ref{sec:calcs} and this section. It transpires that after assignment 
 and analysis of these branches, the values of the experimental 
 $B$ and $D$ constants are reproduced very well, with similar accuracies
 of tens of MHz, by our variational calculations. The planned further
 improvement of the \ai\ PES should lead to the further improvement
 in the agreement between variational and experimental $B$ and $D$ constants
 as well as the agreement between calculated and observed band origins.
 
 What remains to be observed in the HF dimer spectrum
 within the  sensitivity of the 
 existing experimental setup and the Boltzmann factor of the states 
 to be observed is the following:
 $K = 5$ and 6 of the ground vibrational state, $K=1$ and 2 of $\nu_3$
 and $K=1,2,3$ and 4 of $\nu_4$ as well as  $K=0$ of $\nu_6$. These observations could 
 be made using active thermal stabilization of an experimental gas cell and HF generator %certain
 improving the signal-to-noise ratio of the observed spectra  %\red{Tony asks: what improvements to what experiments?} 
 and an expansion of the 
 region of observation to %BWO-24 (range 
 180 - 260 GHz and %BWO-32 (range  
 370 - 535 GHz (accessible with corresponding radiation sources \cite{Istok-site}) %\footnote{\href{http://istokmw.ru/}{istokmw.ru}} 
 in order to observe Q-branches and expand the $J$ ranges
 of the P and R branches. The resulting 
 set of experimental 
 %\textcolor{blue}{$v (is~ not~ constant$)}
  $\Delta_K$, $B$ and $D$ constants would serve as an excellent source of
 data for comparison with an improved \ai\  HF dimer PES.

\section{Discussion and Conclusions}   
\label{sec:concl}

 We present calculations on the 
tunneling-rotation-vibration spectrum of the HF dimer  performed using the variational nuclear motion 
program suite WAVR4 \cite{jt339,newWAVR4} and an \ai\ HF dimer PES \cite{19HuYaZh.HFHF}.
We used the 
results of these calculations to  analyze  a 30-years old gas phase
tunneling-rotation  spectrum of the HF dimer, recorded using a RAD-3
spectrometer in the sub-millimeter region \cite{90BeKaKo.HFHF}. Excellent reproduction
of the observed $B$ constants of the previously assigned \cite{90BeKaKo.HFHF} 
$K_a=0,1,2$ and $3$ branches encouraged us %to try 
to assign further branches of
the  HF dimer %tunneling-rotation
spectrum. 
In particular, the higher \ka\ = 4 branch of the ground vibrational state
 and the two \ka\ = 0 branches of the exited $\nu_4$ and $\nu_3$
 fundamental vibrational inter-molecular modes %tunneling rotation spectra
were assigned for the first time. No previous high resolution 
%spectrum 
study had identified transitions in the $\nu_4$ and $\nu_3$ 
states; we were able to identify lines of these %bands 
states in our sub-millimeter spectrum  on the basis of
 predictions obtained using high accuracy excited $J$ variational calculations.

The present study paves the way for the further improvements of the \ai\ HF dimer
PES and application of it to the accurate calculation of HF dimer
spectrum as well as the assignment of the further energetically
higher lying branches. This should help fully characterise
the HF dimer spectrum  up to dissociation, provided the corresponding
experimental tools are available.
Further improvement of  the \ai\ calculations are necessary
to achieve agreement with experiment.  This work currently is in progress. 

 The  drive for obtaining a more complete understanding of HF dimer
 spectra is not limited to an interest in the HF dimer itself.
 The  HF dimer is similar
 in many respects to the water dimer which enhances the motivation for studying 
 highly excited rovibrational states of (HF)$_2$.
 %Let us consider this similarity in a bit more detail.
 The ability to accurately calculate highly excited  $J$
levels  of the water dimer remains an unsolved
problem but one which is important  for both %theoretical 
fundamental molecular spectroscopy and atmospheric physics.
Developing viable theoretical methods for this  is a prerequisite for the 
calculation of 
accurate water dimer line lists which would provide an important step 
towards a solution
of various physical problems, not the least of which is
the problem of the water continuum in the Earth's atmosphere \cite{12ShPtRa.H2O,17SeOdTr.H2O}
% Shine, K. P., Ptashnik, I.V., Rädel, G., “The Water Vapour Continuum: Brief History and Recent Developments,” Surv Geophys 33(3), 535-555 (2012)
%
%2) E.A. Serov, T.A. Odintsova, M.Yu. Tretyakov, V.E. Semenov, On the origin of the water vapor continuum absorption within rotational and fundamental vibrational bands, J. Quant. Spectrosc. Radiat. Transf. 193 (2017) 1–12. 
%
and
 atmospheres of  exoplanets.
The unambiguous observation  of rotational 
features %in spectra
of water dimers in equilibrium water vapor at close to atmospheric  conditions 
\cite{13TrSeKo.H2O,14SeKoOd.H2O,18KoLeSe.H2O} 
paves the way to a fuller understanding of the  water dimer
spectrum  via a complete theoretical line list and further 
experimental observations, see \cite{19OdTrZi.H2O,20OdTrSi.H2O}.
However, a reliable interpretation of such experimental observations 
requires an accurate theoretical line list, which includes treatment of 
highly excited vibrational and rotational states. This problem 
remains very challenging.

 The HF dimer  system has strong similarities with
 the water dimer: It has the same number of electrons and
 is similarly floppy. The tunneling-rotational spectrum
 of these two dimers are quite similar in terms  of the adiabatic
 relations between the low energy and high energy motions. However, 
 there are also a few significant differences. The water dimer
 is of much higher importance for applications to the atmospheric studies of the
 Earth and exoplanets, and to an
 understanding of liquid water. The second difference is 
 the number of nuclear degrees of freedom: 6 for (HF)$_2$ and 12 for (H$_2$O)$_2$. Because of this
 it is much harder to calculate accurately both the PES and tunneling-rotation-vibration energy levels of 
 (H$_2$O)$_2$, than in the case of (HF)$_2$. These similarities and differences open
 wide opportunities for the study of both dimers. Studies of (HF)$_2$  are of
 increased importance as a similar but easier to study
 system, which will teach us how to deal with the
 more important and more complicated water dimer.
 The third big difference between these two dimers is the relative ease
 of observing  HF dimer spectra  under equilibrium and relatively warm gas phase thermodynamic conditions, 
 as demonstrated by
 numerous studies in both the microwave and the %infrared (
 IR regions, and the utter impossibility
 of observing spectra of similar high resolution 
 for the water dimer.
 This fact makes studies of the HF dimer even more important, as high accuracy high
 resolution experimental data facilitates detailed comparison with theory.
 Such comparisons give guidance on how to achieve comparable accuracy
of calculations for the water dimer, without having to actually
 compare these  calculations with highly excited
 vibrational and rotational   water dimer energy levels. Nevertheless,
 extensive data on the water dimer obtained in beams and jets allows
such comparisons to be made for low-lying rotations and vibrations. These comparisons can
 confirm ideas about the similar accuracy of calculations of
energy levels and line intensities for the water- and HF-dimers; this should allow one 
to extrapolate     our understanding of the accuracy of highly excited
HF dimer calculations to practical calculations of highly excited states of the water dimer. 
% (this is repetition of aforesaid) These considerations increase the importance  of accurate calculations for the HF dimer energy levels and the assignment of higher lying energy levels in the experimental spectrum of this dimer.

The resulting observed minus calculated residuals
for the HF dimer based on experimental and semi-experimental 
data, presented here and the corresponding
variationally calculated values,  provide 
valuable information for  estimating the accuracy of a
corresponding 
water dimer \ai\ calculations of energy levels and line positions.
Clearly this will be the case when
\ai\ calculations on the water dimer are performed at the same level
of electronic structure theory. 
 % (this is repetition of aforesaid) The importance of understanding HF dimer is underscored by the fact that the corresponding water dimer experimental studies cannot readily be performed due to the intrinsic limitations on gas phase high resolution observation of water dimers.The expected similarity in the residues for the two dimers follows directly from the similarity of their electronic structures. 
Thus our HF dimer study provides
useful underpinning information for \ai\ studies on the water dimer supporting our objective,
which  is to calculate a very accurate water dimer \ai\ PES whose accuracy can be
benchmarked by studies of the HF dimer. 

There are two alternative approaches to  further improving our HF dimer PES and hence spectrum calculations.
The first is to produce an improved PES by fitting to experimental
data, as Klopper \etal\ have done \cite{98KlWiQu.HFHF}. However, this approach will 
not guide us to improved water dimer calculations. 
The other alternative is to increase the levels of electronic structure
theory. Work  along these lines is currently in progress  paved by impressive results reported in this paper. 

\section*{Acknowledgments}
We thank Tony Lynas-Gray for helpful comments on our manuscript.
We also thank %the Russian Fund for Fundamental Studies (project 18-02-00705), 
the UK Natural Environment Research Council (NERC) 
grant NE/T000767
and ERC Advanced Investigator Project 883830
for supporting  aspects of this project. %\red{Tony comment: what state?} 
%\textcolor{green}{RF }
RF~State Project 0030-2021-0016 is also acknowledged.

\section*{Supplementary material}
A text files giving the calculated HF dimer energy levels is given  as supplementary data.

%\textcolor{blue}{(see my comments on Refs in that PDF, also check case and position (superscript or subscript) of characters in paper titles: e.g., use HF instead of hf, H$_2$O instead of h2o, etc., etc., etc.)}

\bibliographystyle{elsarticle-num}

%\bibliography{bib/journals_phys,bib/jtj,bib/HFHF,bib/otherpapers,bib/programs}

\begin{thebibliography}{10}
\expandafter\ifx\csname url\endcsname\relax
  \def\url#1{\texttt{#1}}\fi
\expandafter\ifx\csname urlprefix\endcsname\relax\def\urlprefix{URL }\fi
\expandafter\ifx\csname href\endcsname\relax
  \def\href#1#2{#2} \def\path#1{#1}\fi

\bibitem{88Nexxxx.HFHF}
D.~J. Nesbitt, High-resolution infrared spectroscopy of weakly bound molecular
  complexes, Chemical Reviews 88~(6) (1988) 843--870.

\bibitem{74YaNeSc.HFHF}
D.~R. Yarkony, S.~V. O'Neil, H.~F. Schaefer, C.~P. Baskin, C.~F. Bender,
  \href{https://doi.org/10.1063/1.1681161}{Interaction potential between two
  rigid {HF} molecules}, J. Chem. Phys. 60~(3) (1974) 855--865.
\newblock \href {http://arxiv.org/abs/https://doi.org/10.1063/1.1681161}
  {\path{arXiv:https://doi.org/10.1063/1.1681161}}, \href
  {http://dx.doi.org/10.1063/1.1681161} {\path{doi:10.1063/1.1681161}}.
\newline\urlprefix\url{https://doi.org/10.1063/1.1681161}

\bibitem{91QuSuxx.HFHF}
M.~Quack, M.~A. Suhm, Potential energy surfaces, quasiadiabatic channels,
  rovibrational spectra, and intramolecular dynamics of {(HF)$_2$} and its
  isotopomers from quantum monte carlo calculations, J. Chem. Phys. 95~(1)
  (1991) 28--59.

\bibitem{96KlWiQu.HFHF}
W.~Klopper, M.~Quack, M.~A. Suhm, A new ab initio based six-dimensional
  semi-empirical pair interaction potential for {HF}, Chem. Phys. Lett.
  261~(1-2) (1996) 35--44.

\bibitem{98KlWiQu.HFHF}
W.~Klopper, M.~Quack, M.~A. Suhm, {HF} dimer: empirically refined analytical
  potential energy and dipole hypersurfaces from ab initio calculations, J.
  Chem. Phys. 108~(24) (1998) 10096--10115.

\bibitem{19HuYaZh.HFHF}
J.~Huang, D.~Yang, Y.~Zhou, D.~Xie, A new full-dimensional ab initio
  intermolecular potential energy surface and vibrational states for {(HF)$_2$}
  and {(DF)$_2$}, J. Chem. Phys. 150~(15) (2019) 154302.

\bibitem{07BeDeZo.H2O-HF}
S.~P. Belov, V.~M. Demkin, N.~F. Zobov, E.~N. Karyakin, A.~F. Krupnov, I.~N.
  Kozin, O.~L. Polyansky, M.~{\relax Yu}. Tretyakov, Microwave study of the
  submillimeter spectrum of the {H$_2$O$\dots$HF} dimer, J. Mol. Spectrosc.
  241~(2) (2007) 124--135.

\bibitem{87LeMiWi.HCN-HF}
A.~C. Legon, D.~J. Millen, L.~C. Willoughby, Fermi resonance perturbations
  between ($\nu\beta$= n) and ($\nu\sigma$= 1, $\nu\beta$= n- 2) states in the
  rotational spectrum of {HCN$\cdot$HF}, Chem. Phys. Lett. 141~(6) (1987)
  493--498.

\bibitem{96QuSuxx.HFHF}
M.~Quack, M.~A. Suhm, On hydrogen-bonded complexes: the case of {(HF)$_2$},
  Theor. Chim. Acta. 93~(2) (1996) 61--65.

\bibitem{90BeKaKo.HFHF}
S.~P. Belov, E.~N. Karyakin, I.~N. Kozin, A.~F. Krupnov, O.~L. Polyansky,
  M.~{\relax Yu}. Tretyakov, N.~F. Zobov, R.~D. Suenram, W.~J. Lafferty,
  Tunneling-rotation spectrum of the hydrogen fluoride dimer, J. Mol.
  Spectrosc. 141~(2) (1990) 204--222.

\bibitem{72DyHoKl.HFHF}
T.~R. Dyke, B.~J. Howard, W.~Klemperer, Radiofrequency and microwave spectrum
  of the hydrogen fluoride dimer; a nonrigid molecule, J. Chem. Phys. 56~(5)
  (1972) 2442--2454.

\bibitem{83PiLaxx.HFHF}
A.~S. Pine, W.~J. Lafferty, Rotational structure and vibrational
  predissociation in the {HF} stretching bands of the {HF} dimer, J. Chem.
  Phys. 78~(5) (1983) 2154--2162.

\bibitem{87LaSuLo.HFHF}
W.~J. Lafferty, R.~D. Suenram, F.~J. Lovas, Microwave spectra of the
  {(HF)$_2$},{(DF)$_2$}, {HFDF}, and {DFHF} hydrogen-bonded complexes, J. Mol.
  Spectrosc. 123~(2) (1987) 434--452.

\bibitem{85Poxxxx.H2O}
O.~L. Polyansky, One-dimensional approximation of the effective rotational
  hamiltonian of the ground state of the water molecule, J. Mol. Spectrosc.
  112~(1) (1985) 79--87.

\bibitem{87BeKoPo.H2O}
S.~P. Belov, I.~N. Kozin, O.~L. Polyansky, M.~{\relax Yu}. Tret'yakov, N.~F.
  Zobov, Rotational spectrum of the {H$_2 ^{16}$O} molecule in the (010)
  excited vibrational state, J. Mol. Spectrosc. 126~(1) (1987) 113--117.

\bibitem{96PoBuGu.H2O}
O.~L. Polyansky, J.~R. Busler, B.~Guo, K.~Zhang, P.~F. Bernath, The emission
  spectrum of hot water in the region between 370 and 930 \cm, J. Mol.
  Spectrosc. 176~(2) (1996) 305--315.

\bibitem{84PiLaHo.HFHF}
A.~S. Pine, W.~J. Lafferty, B.~J. Howard, Vibrational predissociation,
  tunneling, and rotational saturation in the {HF} and {DF} dimers, J. Chem.
  Phys. 81~(7) (1984) 2939--2950.

\bibitem{95QuSuxx.HFHF}
M.~Quack, M.~A. Suhm, Accurate quantum monte carlo calculations of the
  tunneling splitting in {(HF)$_2$} on a six-dimensional potential
  hypersurface, Chem. Phys. Lett. 234~(1-3) (1995) 71--76.

\bibitem{90QuSuxx.HFHF}
M.~Quack, M.~A. Suhm, Potential energy surface and energy levels of {(HF)$_2$}
  and its {D} isotopomers, Mol. Phys. 69~(4) (1990) 791--801.

\bibitem{90JeBuKa.HFHF}
P.~Jensen, P.~R. Bunker, A.~Karpfen, M.~Kofranek, H.~Lischka, An abinitio
  calculation of the intramolecular stretching spectra for the {HF} dimer and
  its {D}-substituted isotopic species, J. Chem. Phys. 93~(9) (1990)
  6266--6280.

\bibitem{99WuHaMc.HFHF}
X.~T. Wu, E.~F. Hayes, A.~B. McCoy, Rotation--vibration interactions in
  {(HF)$_2$}. {II}. rotation--vibration interactions in low-lying vibrational
  states, J. Chem. Phys. 110~(5) (1999) 2365--2375.

\bibitem{jt339}
I.~N. Kozin, M.~M. Law, J.~Tennyson, J.~M. Hutson, {New vibration-rotation code
  for tetraatomic molecules WAVR4}, Comput. Phys. Commun. 163 (2004) 117--131.

\bibitem{jt312}
I.~N. Kozin, M.~M. Law, J.~M. Hutson, J.~Tennyson, {The rovibrational bound
  states of Ar$_2$HF}, J. Chem. Phys. 118 (2003) 4896--4904.

\bibitem{jt346}
I.~N. Kozin, M.~M. Law, J.~Tennyson, J.~M. Hutson, {Calculating energy levels
  of isomerizing tetraatomic molecules: II. The vibrational states of acetylene
  and vinylidene}, J. Chem. Phys. 122 (2005) 064309.

\bibitem{jt46}
J.~Tennyson, B.~T. Sutcliffe, {Highly rotationally excited states of floppy
  molecules: {H$_2$D$^+$ with J$\le$20}}, Mol. Phys. 58 (1986) 1067--1085.

\bibitem{jt479}
A.~Urru, I.~N. Kozin, G.~Mulas, B.~J. Braams, J.~Tennyson, {Ro-vibrational
  spectra of C$_2$H$_2$ based on variational nuclear motion calcalculations},
  Mol. Phys. 108 (2010) 1973--1990.
\newblock \href {http://dx.doi.org/10.1080/00268976.2010.499858}
  {\path{doi:10.1080/00268976.2010.499858}}.

\bibitem{jt553}
O.~L. Polyansky, I.~N. Kozin, R.~I. Ovsyannikov, P.~Ma\'{l}yszek, J.~Koput,
  J.~Tennyson, S.~N. Yurchenko, {Variational calculation of highly excited
  rovibrational energy levels of H$_2$O$_2$}, J. Phys. Chem. A 117 (2013)
  7367--7377.

\bibitem{newWAVR4}
R.~I. Ovsyannikov, V.~{\relax Yu}. Makhnev, N.~F. Zobov, I.~N. Kozin, O.~L.
  Polyansky, J.~Tennyson, {\sc WAVR4}: a program suite for calculating spectra
  of tetratomic molecules, CPC.

\bibitem{jt338}
J.~Tennyson, M.~A. Kostin, P.~Barletta, G.~J. Harris, O.~L. Polyansky,
  J.~Ramanlal, N.~F. Zobov, {DVR3D: a program suite for the calculation of
  rotation-vibration spectra of triatomic molecules}, Comput. Phys. Commun. 163
  (2004) 85--116.

\bibitem{96AnDaNeb.HFHF}
D.~T. Anderson, S.~Davis, D.~J. Nesbitt, Hydrogen bond spectroscopy in the near
  infrared: Out-of-plane torsion and antigeared bend combination bands in
  {(HF)$_2$}, J. Chem. Phys. 105~(11) (1996) 4488--4503.

\bibitem{96AnDaNea.HFHF}
D.~T. Anderson, S.~Davis, D.~J. Nesbitt, Probing hydrogen bond potentials via
  combination band spectroscopy: A near infrared study of the geared bend/van
  der waals stretch intermolecular modes in {(HF)$_2$}, J. Chem. Phys. 104~(16)
  (1996) 6225--6243.

\bibitem{88PuQuSu.HFHF}
K.~Von~Puttkamer, M.~Quack, M.~A. Suhm, Observation and assignment of
  tunnelling-rotational transitions in the far infrared spectrum of {(HF)$_2$},
  Mol. Phys. 65~(5) (1988) 1025--1045.

\bibitem{90QuSuxb.HFHF}
M.~Quack, M.~A. Suhm, Observation and assignment of the hydrogen bond exchange
  disrotatory in-plane bending vibration $\nu_5$ in {(HF)$_2$}, Chem. Phys.
  Lett. 171~(5-6) (1990) 517--524.

\bibitem{89PuQuSu.HFHF}
K.~von Puttkamer, M.~Quack, M.~A. Suhm, Infrared spectrum and dynamics of the
  hydrogen bonded dimer {(HF)$_2$}, Infrared Physics 29~(2-4) (1989) 535--539.

\bibitem{87PuQuxx.HFHF}
K.~von Puttkamer, M.~Quack, High resolution interferometric ftir spectroscopy
  of {(HF)$_2$}: analysis of a low frequency fundamental near 400 \cm, Mol.
  Phys. 62~(5) (1987) 1047--1064.

\bibitem{Istok-site}
\href{http://istokmw.ru/}{{JSC RPC "Istok" named after Shokin}}, Tech. rep.
\newline\urlprefix\url{http://istokmw.ru/}

\bibitem{12KrTrBe.methods}
A.~F. Krupnov, M.~{\relax Yu}. Tretyakov, S.~P. Belov, G.~{\relax Yu}.
  Golubiatnikov, V.~V. Parshin, M.~A. Koshelev, D.~S. Makarov, E.~A. Serov,
  Accurate broadband rotational {BWO}-based spectroscopy, J. Mol. Spectrosc.
  280 (2012) 110--118.

\bibitem{98BeTrKo.SO2}
S.~P. Belov, M.~{\relax Yu}. Tretyakov, I.~N. Kozin, E.~Klisch, G.~Winnewisser,
  W.~J. Lafferty, J.-M. Flaud, High frequency transitions in the rotational
  spectrum of {SO$_2$}, J. Mol. Spectrosc. 191~(1) (1998) 17--27.

\bibitem{12ShPtRa.H2O}
K.~P. Shine, I.~V. Ptashnik, G.~R{\"a}del, The water vapour continuum: brief
  history and recent developments, Surveys in geophysics 33~(3-4) (2012)
  535--555.

\bibitem{17SeOdTr.H2O}
E.~A. Serov, T.~A. Odintsova, M.~{\relax Yu}. Tretyakov, V.~E. Semenov, On the
  origin of the water vapor continuum absorption within rotational and
  fundamental vibrational bands, J. Quant. Spectrosc. Radiat. Transf. 193
  (2017) 1--12.

\bibitem{13TrSeKo.H2O}
M.~{\relax Yu}. Tretyakov, E.~A. Serov, M.~A. Koshelev, V.~V. Parshin, A.~F.
  Krupnov, Water dimer rotationally resolved millimeter-wave spectrum
  observation at room temperature, Phys. Rev. Lett. 110~(9) (2013) 093001.

\bibitem{14SeKoOd.H2O}
E.~A. Serov, M.~A. Koshelev, T.~A. Odintsova, V.~V. Parshin, M.~{\relax Yu}.
  Tretyakov, Rotationally resolved water dimer spectra in atmospheric air and
  pure water vapour in the 188--258 {GHz} range, Phys. Chem. Chem. Phys.
  16~(47) (2014) 26221--26233.

\bibitem{18KoLeSe.H2O}
M.~A. Koshelev, I.~I. Leonov, E.~A. Serov, A.~I. Chernova, A.~A. Balashov,
  G.~M. Bubnov, A.~F. Andriyanov, A.~P. Shkaev, V.~V. Parshin, A.~F. Krupnov,
  et~al., New frontiers in modern resonator spectroscopy, IEEE Transactions on
  Terahertz Science and Technology 8~(6) (2018) 773--783.

\bibitem{19OdTrZi.H2O}
T.~A. Odintsova, M.~{\relax Yu}. Tretyakov, A.~O. Zibarova, O.~Pirali, P.~Roy,
  A.~Campargue, Far-infrared self-continuum absorption of {H$_2^{16}$O} and
  {H$_2^{18}$O} (15--500 \cm), J. Quant. Spectrosc. Radiat. Transf. 227 (2019)
  190--200.

\bibitem{20OdTrSi.H2O}
T.~A. Odintsova, M.~{\relax Yu}. Tretyakov, A.~A. Simonova, I.~V. Ptashnik,
  O.~Pirali, A.~Campargue, Measurement and temperature dependence of the water
  vapor self-continuum between 70 and 700 \cm, Journal of Molecular Structure
  (2020) 128046.

\end{thebibliography}

\end{document}